\documentclass[12pt, a4paper, reqno]{amsart}

\usepackage{amssymb}
\usepackage[left=3.3cm, right=3.3cm, top=3.5cm, bottom=3.5cm]{geometry}
\usepackage{microtype}

\theoremstyle{plain}
\newtheorem{thm}{Theorem}

\theoremstyle{remark}

\newcommand{\stirling}{\genfrac{\{}{\}}{0pt}{}}

\begin{document}

\title{A note on multipivot Quicksort}

\author{Vasileios Iliopoulos}

\address{Department of Mathematical Sciences, 
University of Essex, Colchester, UK} 

\email{viliop@essex.ac.uk, iliopou@gmail.com}

\date{\today}



\begin{abstract}
We analyse a generalisation of the Quicksort algorithm, where $k$ uniformly
at random chosen pivots are
used for partitioning an array of $n$ distinct keys. 
Specifically, the expected cost of
this scheme is obtained, under the assumption of linearity of
the cost needed for the partition process. The integration constants of the 
expected cost are computed using Vandermonde matrices.
\end{abstract}

\maketitle

\section{Introduction}

The Quicksort algorithm invented by Hoare \cite{hoare} sorts $n$ keys 
by randomly choosing a key called pivot and rearranging the array by
comparing every key to the pivot, so that
all keys less than or equal to the pivot are on its left 
and all keys greater than or equal to the pivot are on its right.
The algorithm is then recursively applied to each of these two smaller arrays (which either might be
empty) till we get trivial arrays of
length $1$ or $0$. The term ``key'' can be a number, word and more generally can be an element of 
a finite set, equipped with a transitive relation.
Throughout this note, we assume
that the input array is a random permutation of the positive integers $\{1, \ldots, n\}$ with all the
$n!$ permutations equally likely to be the input.

A generalisation of the algorithm is to randomly choose
$k$ pivots $i_{1}, i_{2}, \ldots, i_{k}$, where $k = 1, 2, \ldots $ and
partition the array to $(k+1)$ subarrays. 
The algorithm is recursively applied to each
of the segments that contains at least $(k+1)$ keys and arrays with less than
$(k+1)$ keys are sorted by another algorithm, as insertion sort. 
We point out at once that this multipivot Quicksort is a
special case of Hennequin's `generalised Quicksort' \cite{Hen}, where a
random sample of $k(t+1)-1$ keys is chosen from the array
to be sorted and the $(t+1)$-st, $2(t+1)$-th, \ldots,
$(k-1)(t+1)$-th smallest keys are used as pivots.
Obviously, for $t=0$, the array is partitioned to $k$ subarrays, 
according to $k-1$ pivots. For 
$k=2$, we have the `median of $(2t+1)$' Quicksort. For other multipivot variants, we
also refer the reader to the Ph.D. theses of Sedgewick \cite{sedg} and Tan \cite{tan}.

In this note, we consider the average case
analysis of multipivot Quicksort and compute the constants of integration by Vandermonde
matrices. Let $f(n, k)$ denote the expected cost when a randomly permuted array of $n$ keys is to 
be sorted by the application of Quicksort on $k$ pivots. 
We deliberately allow some flexibility in the form of the cost, 
but a typical example might be the number of
comparisons made. We obtain the following recursive relation:
\begin{align*}
f(n, k) & = T(n, k)  \\
& \quad {}  + \dfrac{1}{\dbinom{n}{k}}  \underbrace{\sum_{i'_{1}} \sum_{i'_{2}} \ldots 
\sum_{i'_{k}}}_{i'_{1} < i'_{2} < \ldots < i'_{k}} 
\Bigl ( f(i'_{1}-1, k) + f(i'_{2} - i'_{1} -1, k) 
+ \ldots + f(n - i'_{k}, k ) \Bigr),
\end{align*}
where $\mathbb{E}(\tau(n, k)) = T(n, k) = \overline{a}(k)n + \overline{b}(k)$ is the average value 
of a ``toll function'' $\tau(n, k)$ during the first partitioning stage. 
We assume that this is a linear function of $n$. 
The recursion may look a complex $k$-index summation, 
but can be simplified by noting the ranges of the indices;
\begin{align*}
f(n, k) 
& = T(n, k) +  \dfrac{1}{\dbinom{n}{k}} \sum_{i'_{1}=1}^{n - k+1} \sum_{i'_{2}= 
i'_{1}+1}^{n-k+2} \ldots \sum_{i'_{k} = i'_{k-1} + 1}^{n} \Bigl ( f(i'_{1}-1, k) + \ldots + f(n - i'_{k}, k ) \Bigr ) \\
& = T(n, k) + \dfrac{(k+1)!}{n(n-1) \ldots (n-k+1)}
\sum_{i'_{1} = 1}^{n-k+1} \dbinom{n - i'_{1}}{k-1}f(i'_{1}-1, k),
\end{align*}
since the partitioning of the array according to $k$ pivots yields $(k+1)$
segments and using the fact that the expectations of the average costs in
each segment are equal owing to the uniform distribution of the permutation.

\section{Solution of the Cauchy-Euler differential equation}

With the view of applying generating functions for the solution of our recurrence, 
let $f(n, k) = a_{n}$ and consider $h(x)= \displaystyle \sum_{n=0}^{\infty}a_{n}x^{n}$;
\begin{eqnarray*}
\sum_{n=0}^{\infty}\dbinom{n}{k}a_{n}x^{n} = 
\sum_{n=0}^{\infty} \dbinom{n}{k}T(n, k)x^{n} + (k+1)\sum_{n=0}^{\infty} 
\left (\sum_{i'_{1} = 1}^{n} \dbinom{n - i'_{1}}{k-1}a_{i'_{1}-1} \right )x^{n}.
\end{eqnarray*}
Interchanging the order of summation and multiplying both sides 
by $\left (\dfrac{x}{1-x} \right )^{-k}$, 
this becomes a $k$-th order differential equation
\begin{eqnarray*}
(1-x)^{k}h^{(k)}(x) = \dfrac{k!\bigl (\overline{a}(k)(x+k)+ 
\overline{b}(k)(1-x) \bigr )}{(1-x)^{2}} + h(x)(k+1)!, 
\end{eqnarray*}
which is a {\em Cauchy--Euler} differential equation. This type of differential
equations is inherent to the analysis of Quicksort and its variants:  
we refer the reader to \cite{hc}, \cite{dur}, \cite{Hen, Hennequin} and \cite{sedg}. 
Substituting $z=1-x$, we have $h(x)=g(1-x)$ and
\begin{eqnarray*}
(-1)^{k}z^{k}g^{(k)}(z)- g(z)(k+1)! = 
\dfrac{k!\left (\overline{a}(k)(1-z+k)+ \overline{b}(k)z \right )}{z^{2}}.
\end{eqnarray*}

Following the analysis of Hennequin \cite{Hen, Hennequin} and Sedgewick \cite{sedg}, we 
use the differential operator $\Theta$, with $\Theta g(z):=zg^{\prime}(z)$ for the
solution of the differential equation. It is easily verifiable by
induction that $\binom{\Theta}{k}g(z) =
\frac{z^{k}g^{(k)}(z)}{k!}$ and we have
\begin{eqnarray*}
\bigl ((-1)^{k}\Theta(\Theta -1) \ldots (\Theta -k + 1)-(k+1)! \bigr )g(z) = 
\dfrac{k!\left (\overline{a}(k)(1-z+k)+ \overline{b}(k)z \right )}{z^{2}}.
\end{eqnarray*}
The indicial polynomial $\mathcal{P}_{k}(\Theta)$ is equal to
\begin{equation*}
\mathcal{P}_{k}(\Theta)=(-1)^{k}\Theta^{\underline{k}}-(k+1)!,
\end{equation*}
where using the notation from \cite{Gra}, 
$\Theta^{\underline{k}}:=\Theta(\Theta-1)\cdot \ldots \cdot (\Theta-k+1)$ 
denotes the falling factorial.

It can be easily proved that the polynomial has $k$ simple roots, with real parts in $[-2, k+1]$. 
The solution of the differential equation is
\begin{align}
g(z) & = \dfrac{\overline{a}(k)(k+1)!}{(-2-r_{1})(-2-r_{2}) 
\ldots (-2-r_{k-1})}\dfrac{\ln(z)}{z^{2}} \notag \\
& \qquad {} + \dfrac{k!}{(-1-r_{1})(-1-r_{2}) \ldots 1}\dfrac{\bigl 
(\overline{b}(k)-\overline{a}(k) \bigr )}{z} 
+ \sum_{i=1}^{k}s_{i}z^{r_{i}}.
\label{1}
\end{align}
In order to evaluate $\mathcal{S}_{k-1}(-2)=(-2-r_{1})(-2-r_{2}) 
\ldots (-2-r_{k-1})$, note that
\begin{equation*}
\mathcal{S}_{k-1}(-2) = \mathcal{P}_{k}^{\prime}(-2),
\end{equation*}
thus
\begin{equation*}
\mathcal{S}_{k-1}(-2) = -(k+1)!(H_{k+1} - 1).
\end{equation*}
Moreover,
\begin{equation*}
\mathcal{P}_{k}(-1)= -kk!
\end{equation*}

and in terms of series, we have
\begin{align}
h(x) & = \dfrac{\overline{a}(k)}{H_{k+1} - 1} \sum_{n=0}^{\infty}
\bigl ( (n+1)H_{n}-n \bigr ) x^{n} +
\sum_{n=0}^{\infty} \sum_{i=1}^{k}s_{i}(-1)^{n}\dbinom{r_{i}}{n}x^{n} \notag \\
& \qquad {} +
\dfrac{\overline{a}(k)-\overline{b}(k)}{k} \sum_{n=0}^{\infty}x^{n}.
\label{2}
\end{align}
Extracting the coefficients and noting that $-2$ is the unique root with
the least real part, the expected cost of Quicksort on $k$ uniformly at random
chosen pivots is 
\begin{align*}
a_{n} & = \dfrac{\overline{a}(k)}{H_{k+1} - 1} \bigl ((n+1)H_{n}-n \bigr )
+ s_{k}(n+1)+o(n).
\end{align*}

\section{Computation of the integration constants using Vandermonde matrices}

In this section, we compute the constants of integration $s_{i}$ using Vandermonde
determinants. We remark that this approach is employed in \cite{dur},
where the nine integration constants involved in the expected number
of comparisons of `remedian of $3^{2}$' Quicksort
are computed using Vandermonde matrices.
In \cite{Hen, Hennequin}, the constant
corresponding to the root $-2$ is computed by the application of generating 
functions and the differential
operator (see Proposition {\bf III.8} in \cite[page 50]{Hen}). Also,
Vandermonde determinants appear in the analysis of multiple Quickselect \cite{panh}.
Our system of equations is
\begin{eqnarray*}
g(1)=g^{\prime}(1)= \ldots = g^{(k-1)}(1) = 0.
\end{eqnarray*}
Differentiating $m$ times Eq. \eqref{1} and setting $z=1$,
\begin{eqnarray}
\sum_{i=1}^{k}s_{i}r_{i}^{\underline{m}}=(-1)^{m+1}m! \biggl 
(\dfrac{\overline{a}(k)\bigl(m+1)H_{m}-m)
 \bigr)}{(H_{k+1}-1)}+ \dfrac{\overline{a}(k)-\overline{b}(k)}
{k} \biggr ),
\label{3}
\end{eqnarray}
for $m=0, 1, \ldots, k-1.$
In matrix form, Eq. \eqref{3} is
\begin{eqnarray*}
\centering
&& \begin{bmatrix}
1 & 1 & \ldots & 1 \\
r_{1} & r_{2} & \ldots & -2  \\
\vdots & \vdots & \ddots & \vdots  \\
r^{\underline{k-1}}_{1} & r^{\underline{k-1}}_{2} & \ldots & (-2)^{\underline{k-1}} \\
\end{bmatrix}
\begin{bmatrix}
s_{1} \\
s_{2} \\
\vdots \\
s_{k}
\end{bmatrix} = \\
\centering
&& \begin{bmatrix}
& -\dfrac{1}{k}\bigl (\overline{a}(k)-\overline{b}(k) \bigr ) \\
& \dfrac{\overline{a}(k)}{H_{k+1} - 1} + \dfrac{1}{k}\bigl (\overline{a}(k)-\overline{b}(k) \bigr ) \\
& \vdots \\
& (-1)^{k}(k-1)!\biggl 
(\dfrac{\overline{a}(k)\bigl (kH_{k-1}-(k-1)
\bigr)}{H_{k+1}-1} + \dfrac{\overline{a}(k)-\overline{b}(k)}
{k} \biggr )
\end{bmatrix}
\end{eqnarray*}
It is easy to see that the coefficient matrix is non-singular. 
Using the generating function $x^{k-1}=\sum_{j=1}^{\infty} \stirling{k-1}{j-1}x^{\underline{j-1}}$ \cite{Abr},
where $\stirling{n}{k}$ are the Stirling numbers of the second kind, with $\stirling{n}{k}=0$ for $n<k$, 
we can write each power 
of $r_{i}$ as a sum of integer multiples of (earlier) 
rows of the matrix of coefficients we get naturally (and that of course does not change the determinant). 
Hence the determinant of this matrix is the same as the determinant of the Vandermonde matrix, 
which is well known to be $\prod_{1 \leq i < j \leq k} (r_{j}-r_{i}) \neq 0$, as the roots are all simple. 

Transforming the matrix into a Vandermonde one, we have
\begin{eqnarray*}
\sum_{j=1}^{\infty}(-1)^{j}(j-1)! \stirling{k-1}{j-1} \biggl (
\dfrac{\overline{a}(k)\bigl (jH_{j}-j
)}{H_{k+1}-1} + \dfrac{\overline{a}(k)-\overline{b}(k)}{k} \biggr ).
\end{eqnarray*}
Note that \cite{Abr},
\begin{equation*}
\sum_{j=1}^{k}(-1)^{j}(j-1)! \stirling{k-1}{j-1}=(-1)^{k},
\end{equation*}
since $(-1)^{j-1}(j-1)!=(-1)^{\underline{j-1}}$. Also,
\begin{equation*}
(-1)^{\underline{j}} = (-1)(-2)^{\underline{j-1}},
\end{equation*}
hence
\begin{equation*}
\sum_{j=1}^{k}(-1)^{j}j! \stirling{k-1}{j-1}=(-1)^{k}2^{k-1}.
\end{equation*}
Differentiating the generating function, we have
\begin{equation*}
(k-1)x^{k-2}=\sum_{j=2}^{k} \stirling{k-1}{j-1} x^{\underline{j-1}}\left 
(\sum_{i=0}^{j-2}\dfrac{1}{x-i} \right ),
\end{equation*}
therefore
\begin{align}
(-1)^{k}(k-1)2^{k-2}& = \sum_{j=2}^{k}(-1)^{j}j! \stirling{k-1}{j-1} \left 
(\sum_{i=0}^{j-2}\dfrac{1}{i+2} \right)\notag \\
& = \sum_{j=2}^{k}(-1)^{j}j! \stirling{k-1}{j-1} (H_{j}-1).
\label{4}
\end{align}
Note that Eq. \eqref{4} is a special case of Eq. (36) from \cite{panh}, for $x=1$.

Our linear system now becomes:
\begin{eqnarray*}
\centering
&& \begin{bmatrix}
1 & 1 & \ldots & 1 \\
r_{1} & r_{2} & \ldots & -2  \\
\vdots & \vdots & \ddots & \vdots  \\
r^{k-1}_{1} & r^{k-1}_{2} & \ldots & (-2)^{k-1} \\
\end{bmatrix}
\begin{bmatrix}
s_{1} \\
s_{2} \\
\vdots \\
s_{k}
\end{bmatrix} = \\
\centering
&& \begin{bmatrix}
& -\dfrac{1}{k}\bigl (\overline{a}(k)-\overline{b}(k) \bigr ) \\
& \dfrac{\overline{a}(k)}{H_{k+1} - 1} + \dfrac{1}{k}\bigl (\overline{a}(k)-\overline{b}(k) \bigr ) \\
& \vdots \\
& (-1)^{k} \left ((k-1)2^{k-2}
\dfrac{\overline{a}(k)}{H_{k+1}-1} + \dfrac{\overline{a}(k)-\overline{b}(k)}
{k}  \right )
\end{bmatrix}
\end{eqnarray*}
The inverse matrix can be factored into a product of an upper and lower 
triangular matrices, \cite{Hou}, \cite{Turner}. 
In \cite{Hou} an algorithm is presented, where the entries 
of the triangular matrices are recursively computed.
Letting ${\bf A}^{-1}$ denote the inverse, we have 
\begin{eqnarray*}
{\bf A}^{-1} = 
\begin{bmatrix}
1 & \frac{1}{r_{1}-r_{2}} & \frac{1}{(r_{1}-r_{2})(r_{1}-r_{3})} & \ldots \\
0 &  \frac{1}{r_{2}-r_{1}} & \frac{1}{(r_{2}-r_{1})(r_{2}-r_{3})} & \ldots \\
0 & 0 & \frac{1}{(r_{3}-r_{1})(r_{3}-r_{2})} & \ldots \\
0 & 0 & 0 & \ldots \\
\vdots & \vdots & \vdots & \ldots  
\end{bmatrix}
\begin{bmatrix}
1 & 0 & 0 & \ldots  \\
-r_{1} & 1 & 0 & \ldots  \\
r_{1}r_{2} & -(r_{1}+r_{2}) & 1 & \ldots \\
-r_{1}r_{2}r_{3} & r_{1}r_{2}+r_{1}r_{3}+r_{2}r_{3} & -(r_{1}+r_{2}+r_{3}) & \ldots \\
\vdots & \vdots & \vdots & \ldots  
\end{bmatrix}
\end{eqnarray*}

The constants of integration are given by:
\begin{eqnarray*}
s_{i} =(-1)^{k}\left (\dfrac{\displaystyle \prod_{j \neq i}^{k}
(r_{j}+1)}{\displaystyle \prod_{j \neq  i}^{k}(r_{i} - r_{j})}
\biggl( \dfrac{\overline{a}(k)-\overline{b}(k)}{k} \biggr)+ \dfrac{\overline{a}(k)}{H_{k+1}-1} 
\dfrac{
\displaystyle \sum_{j=1}^{k-1}\biggl(\prod_{l \neq j}^{k-1}(r_{l}+2)\biggr)} 
{\displaystyle \prod_{j \neq i}^{k}(r_{i} - r_{j})}
 \right ).
\end{eqnarray*}
The products of pairwise differences of roots $r_{i}$ and $r_{j}$ that naturally arise in {\bf LU} triangular decomposition of 
the inverse of Vandermonde matrix form alternating polynomial functions. Putting $i=k$ to the
previous equation, 
\begin{equation*}
\dfrac{\displaystyle \prod_{j=1}^{k-1}(r_{j}+1)}{\displaystyle \prod_{j=1}^{k-1}(-2-r_{j})}
=(-1)^{k-1}\dfrac{\mathcal{P}_{k}(-1)}{\mathcal{S}_{k-1}(-2)}=(-1)^{k-1}\dfrac{k}{(k+1)(H_{k+1}-1)}
\end{equation*}
and the sum of the products is
\begin{equation*}
\displaystyle \sum_{j=1}^{k-1}\biggl(\prod_{l \neq j}^{k-1}(r_{l}+2)\biggr)=\mathcal{S}^{\prime}_{k-1}(-2).
\end{equation*} 
Differentiating $\mathcal{P}_{k}(\Theta)$ twice and setting $\Theta=-2$,
\begin{equation*}
\mathcal{S}^{\prime}_{k-1}(-2)=\dfrac{\mathcal{P}_{k}^{\prime \prime}(-2)}{2}=\dfrac{(k+1)!(H^{2}_{k+1}-2H_{k+1}
-H^{(2)}_{k+1}+2)}{2},
\end{equation*}
where $H^{(2)}_{k+1}:=\displaystyle \sum_{j=1}^{k+1}\frac{1}{j^{2}}$ denotes the second--order
harmonic number.

The main result of this paper is the following Theorem:
\begin{thm}
The expected cost of multipivot Quicksort on $k$ uniformly at random selected pivots 
for partitioning an array consisting of $n > k$ distinct keys to
subarrays that each one contains at most $k$ keys is
\begin{align*}
& \dfrac{\overline{a}(k)}{H_{k+1} - 1} \bigl ((n+1)H_{n}-n \bigr ) 
- \Biggl (\dfrac{\overline{a}(k)}{H_{k+1}-1} \biggl
(\dfrac{H^{2}_{k+1}-2H_{k+1}-H^{(2)}_{k+1}+2}{2(H_{k+1}-1)} \biggr)  \\
& \qquad{}+ \dfrac{\overline{a}(k)-\overline{b}(k)}{(k+1)(H_{k+1}-1)} \Biggr )(n+1) 
+ o(n),
\end{align*}
where the ``toll function'' has the average value $\overline{a}(k)n+\overline{b}(k)$.
\end{thm}

\subsubsection*{Acknowledgements}

I thank Dr. David B. Penman for his advice and helpful suggestions regarding this paper.


\begin{thebibliography}{40}

\bibitem{Abr} M. Abramowitz and I. A. Stegun, {\em ``Handbook of Mathematical 
Functions with Formulas, Graphs, and Mathematical Tables''.} Dover Publications, 1972.


\bibitem{hc} H. H. Chern, H. K.  Hwang and T. H. Tsai, 
{\em ``An asymptotic theory for Cauchy--Euler differential equations 
with applications to the analysis of algorithms''.}
J. Algorithm. {\bf 44}, 177-225, 2002.


\bibitem{dur} M. Durand, {\em ``Asymptotic analysis of an optimized Quicksort
algorithm''.} Inform. Proc. Lett. {\bf 85} (2): 73-77, 2003.


\bibitem{Gra}  R. Graham, D. E.  Knuth and O. Patashnik, 
{\em ``Concrete Mathematics: A Foundation for Computer Science''.} 
Addison--Wesley Publishing, 2nd Edition, 1994.



\bibitem{Hen} P. Hennequin, {\em ``Analyse en moyenne d'algorithmes: tri rapide et arbres de recherche''.} 
Ph.D. thesis. \'{E}cole Polytechnique, 1991.


\bibitem{Hennequin} P. Hennequin,  {\em ``Combinatorial analysis of quicksort algorithm''.} 
RAIRO Theor. Inform. Appl. {\bf 23} (3): 317-333, 1989.


\bibitem{hoare}  C. A. R. Hoare, {\em ``Quicksort''.}
Comput. J. {\bf 5} (1): 10-15, 1962.


\bibitem{Hou} S. H. Hou and E. Hou, {\em ``Triangular Factors of the Inverse of Vandermonde Matrices''.} 
Proceedings of the International MultiConference of Engineers 
and Computer Scientists. Vol. II IMECS, Hong Kong, 2008.


\bibitem{panh} A. Panholzer, {\em ``Analysis of multiple quickselect variants''.}
Theor. Comput. Sci. {\bf 302} (1-3): 45-91, 2003.


\bibitem{sedg} R. Sedgewick, {\em ``Quicksort''.} Ph.D. thesis. 
Garland Publishing, 1980.


\bibitem{tan}  K. H. Tan, {\em ``An asymptotic analysis of the number
of comparisons in multipartition quicksort''.} Ph.D. thesis. Carnegie Mellon University,
1993.


\bibitem{Turner} L. R. Turner, {\em ``Inverse of the Vandermonde Matrix with Applications''.} 
National Aeronautics and Space Administration, technical note D-3547, 1966. 

\end{thebibliography}
\end{document}